\documentclass[aps,amsmath,amssymb,reprint,10pt,twocolumn,superscriptaddress]{revtex4-1}
\usepackage{bbm}
\usepackage{graphicx}
\usepackage[countmax]{subfloat}
\usepackage{framed}
\usepackage{color}
\usepackage{hyperref}
\usepackage{epstopdf}
\usepackage{dsfont}
\usepackage{amsthm}
\usepackage{wrapfig}
\usepackage{relsize}
\usepackage{bm}
\usepackage[latin1]{inputenc}
\usepackage{enumitem}
\usepackage{soul}

\newcommand{\red}[1]{{\color{red}}}

\usepackage{pifont}

\newcommand{\ba}{\begin{eqnarray}}
\newcommand{\ea}{\end{eqnarray}}
\begin{document}

\title{Robust calibration of multiparameter sensors via machine learning at the single-photon level} 

\author{Valeria Cimini}
\thanks{These three authors contributed equally}
\affiliation{Dipartimento di Scienze, Universit\`{a} degli Studi Roma Tre, Via della Vasca Navale 84, 00146, Rome, Italy}
\author{Emanuele Polino}
\thanks{These three authors contributed equally}
\affiliation{Dipartimento di Fisica, Sapienza Universit\`{a} di Roma, Piazzale Aldo Moro 5, I-00185 Roma, Italy}
\author{Mauro Valeri}
\thanks{These three authors contributed equally}
\affiliation{Dipartimento di Fisica, Sapienza Universit\`{a} di Roma, Piazzale Aldo Moro 5, I-00185 Roma, Italy}
\author{Ilaria Gianani}
\affiliation{Dipartimento di Scienze, Universit\'a degli Studi Roma Tre, Via della Vasca Navale 84, 00146, Rome, Italy}
\author{Nicol\`o Spagnolo}
\affiliation{Dipartimento di Fisica, Sapienza Universit\`{a} di Roma, Piazzale Aldo Moro 5, I-00185 Roma, Italy}
\author{Giacomo Corrielli}
\affiliation{Istituto di Fotonica e Nanotecnologie, Consiglio Nazionale delle Ricerche (IFN-CNR), Piazza Leonardo da Vinci, 32, I-20133 Milano, Italy}
\affiliation{Dipartimento di Fisica, Politecnico di Milano, Piazza Leonardo da Vinci, 32, I-20133 Milano, Italy}
\author{Andrea Crespi}
\affiliation{Dipartimento di Fisica, Politecnico di Milano, Piazza Leonardo da Vinci, 32, I-20133 Milano, Italy}
\affiliation{Istituto di Fotonica e Nanotecnologie, Consiglio Nazionale delle Ricerche (IFN-CNR), Piazza Leonardo da Vinci, 32, I-20133 Milano, Italy}
\author{Roberto Osellame}
\affiliation{Istituto di Fotonica e Nanotecnologie, Consiglio Nazionale delle Ricerche (IFN-CNR), Piazza Leonardo da Vinci, 32, I-20133 Milano, Italy}
\affiliation{Dipartimento di Fisica, Politecnico di Milano, Piazza Leonardo da Vinci, 32, I-20133 Milano, Italy}
\author{Marco Barbieri}
\affiliation{Dipartimento di Scienze, Universit\'a degli Studi Roma Tre, Via della Vasca Navale 84, 00146, Rome, Italy}
\affiliation{Istituto Nazionale di Ottica - CNR, Largo Enrico Fermi 6, 50125 Florence, Italy}
\author{Fabio Sciarrino}
\affiliation{Dipartimento di Fisica, Sapienza Universit\`{a} di Roma, Piazzale Aldo Moro 5, I-00185 Roma, Italy}

\begin{abstract} 
Calibration of sensors is a fundamental step to validate their operation. This can be a demanding task, as it relies on acquiring a detailed modelling of the device, aggravated by its possible dependence upon multiple parameters. Machine learning provides a handy solution to this issue, operating a mapping between the parameters and the device response, without needing additional specific information on its functioning. Here we demonstrate the application of a Neural Network based algorithm for the calibration of integrated photonic devices depending on two parameters. We show that a reliable characterization is achievable by carefully selecting an appropriate network training strategy. These results show the viability of this approach as an effective tool for the multiparameter calibration of sensors characterized by complex transduction functions.
\end{abstract}

\maketitle

\section{Introduction}

Quantum metrology has demonstrated significant advances in the last few years, keeping up with the perspective of a novel generation of quantum sensors with enhanced sensitivity \cite{paris2009quantum,giovannetti2011advances,pezze2014,pirandola2018advances,polino2020photonic}. In order to fully exploit these advantages, the device must be known and controlled, so that the measured parameters can be retrieved with good accuracy. This ability then relies on the availability of a trusted calibration of the sensor in hand~\cite{Gianani2020assessing}. Conventional methods for device characterisation require, in general, a large set of calibration data and intensive post-processing. Alternatively, one could rely on refined a-priori physical models of the sensors, including noise effects, based on measured quantities.  Such intensive approaches, however,  become unpractical for sensors of increasing complexity, and unfeasible in the perspective of commercial devices.

Further, in realistic sensors, the user has typically access and control over a set of physical parameters, which in turn modify the internal characteristics of the device. Therefore, the target parameters are not directly evaluated, but inferred from measured quantities. Optical phases, determined by electrical signals via thermo-optic effects, are a case in point \cite{flamini2015lsa,carolan2015universal,harris2017transport,taballione2019programmable}: here, voltages are the parameters of interest, but the measured optical signal derives from variations of optical phases. Therefore, the sensor inherently works as a transductor,  mapping the parameters to be estimated onto measured quantities through a suitable response function, which needs to be characterized as well. In this respect, spurious effects also concur to determining the response function, and must be taken into account. This poses major difficulties in the modelling, and increases the complexity of an experimental characterisation via conventional methods. 
 
A practical approach thus requires a different methodology for sensor calibration. A viable direction is provided by machine learning techniques, capable of handling large datasets and of solving tasks for which they have not been explicitly programmed; applications range from stock prices predictions \cite{TICKNOR20135501,ENKE2011201} to the analysis of medical diseases \cite{Ganesan}. In the last few years, several applications of machine learning methods in the quantum domain have been reported \cite{dunjko2018machine,mehta2019high,carleo2019machine}, including state and unitary tomography \cite{spagnolo2017genetic,carrasquilla2019reconstructing,palmieri2020experimental,rocchetto2019,arrazola2019machine,Giordani_2020,Neugebauer2020,PhysRevLett.123.230504,Tiunov:s}, design of quantum experiments \cite{nichols2019designing,melnikov,krenn2016automated,o2019hybrid,sabapathy2019production,krenn2020computer,gao2020computer}, validation of quantum technology \cite{agresti2019prx,flamini2019qst,Knott_2016}, identification of quantum features \cite{Wigner,PhysRevResearch.2.023150}, and the adaptive control of quantum devices \cite{hentschel2010machine, hentschel2011efficient,lovett2013differential,bonato2016optimized,palittapongarnpim2017learning,liu2017control,Paesani2017,piccoloLume,palittapongarnpim2019robustness,dinani2019bayesian,liu2020repetitive,peng2020feedback,rambhatla2020adaptive,valeri2020experimental,craigie2020resource,nolan2020machine,fiderer2020}. Also, photonic platforms can be exploited for the realization of machine learning protocols \cite{bernstein2020freely,flamini2020photonic}. Recently, a first insight on the application of machine learning methods for the calibration of a quantum sensor has been reported \cite{cimini2019calibration}. In detail, the characterization of an optical phase sensor was carried out by means of artificial neural networks (NNs)~\cite{NNbook}. This has demonstrated its advantages, in that it required no detailed model, it relied on the same states for the calibration as for the estimation, and demonstrated robustness to finite-size datasets when compared to standard methods. When extending the use of NNs to multiple parameter scenarios, it can be expected that all these features can be preserved. However, conjugating this approach with the peculiarities of multiparameter estimation is far from obvious. 

In this article, we demonstrate the calibration of multiphase sensors implemented in a femtosecond-laser-written photonic platform. The devices are multiarm interferometers with multiple embedded phases, which are controlled externally by applying a voltage to resistors placed within the sensor. We report on how to build and train a NN to make it capable to work as a reliable and practical calibration tool. Indeed, we show the requirements on the training process, in terms of the size of the dataset, as well as the need to resolve possible ambiguities which result from the non-monotonicity of the sensor response function. Our results give evidence on the viability of a machine learning approach for calibration of complex quantum devices.

\begin{figure*}[htb!]
\begin{center}
{\includegraphics[width=\textwidth]{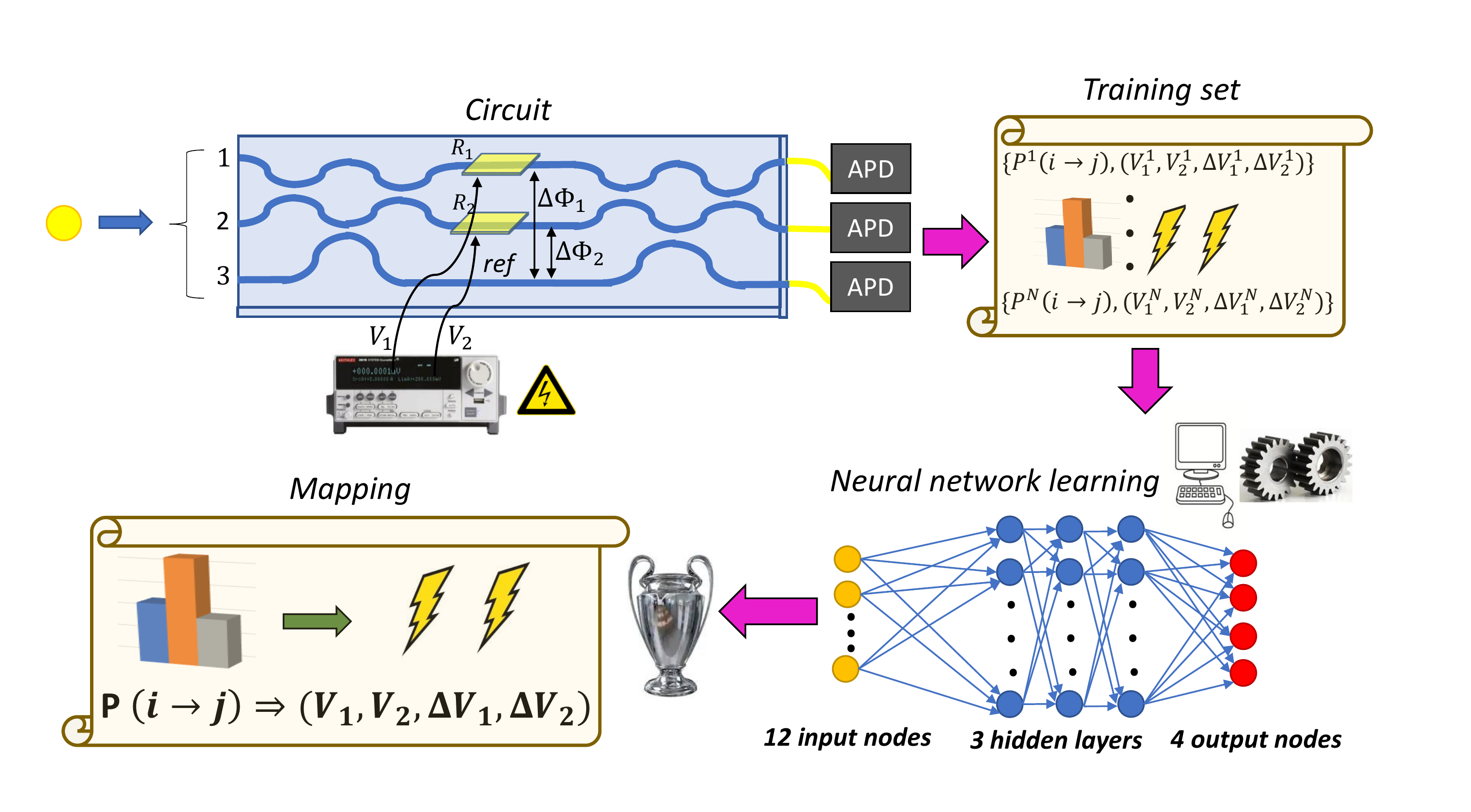}}
\caption{Conceptual scheme, showing the calibration steps in clockwise sense. Single photons are sent along the inputs of the three-arm interferometer and revealed by avalanche photodiodes (APDs). The output probabilities  $P(i\rightarrow j)$ (with $i,j=1,2,3$) are measured as function of the two applied voltages $V_1$ and $V_2$. A small portion of the data set, is used to train the neural network. After the training the NN is able to map any output probability to the corresponding pair of voltages.}
\label{concim}
\end{center}
\end{figure*}

\section{results}
\label{risultati}
 In this section, we detail our use of a feed-forward neural network \cite{Aggarwal18} to approximate the multivariate function that links single-photon detection probabilities, relative to  outputs of an integrated three-mode device, given a certain input, to the voltage settings controlling the interferometer phases.
 
Regression is one of the most common problems faced by supervised learning algorithms. It consists in finding a map $f$ that links an input vector of features $\vec{x}_i$ to the respective output vector $\vec{y}_i$ consisting of real numbers, for all the examples $i$ in the dataset. NNs are very effective at modelling complex non-linear functions for very large dimension dataset described by many features, and their performances are particularly good when large training dataset, containing many different examples, are available.
%Here, \textcolor{blue}{we apply such techniques for the calibration of a device able to perform quantum information tasks with single photons. In particular,} 

\subsection{Experimental platform}

The integrated device under study is a three-arm interferometer realized by the femtosecond laser writing technique  \cite{gattass2008flw,dellavalle2008flw} and able to perform multiphase estimation protocols \cite{polino2019experimental,valeri2020experimental}. The circuit is composed by a sequence of two three-arm beam splitters (tritters) realized through  a two-dimensional geometry decomposition and interposed by three internal arms encoding two independent optical phase shifts $\bm{\Delta \phi} = (\Delta \phi_1, \Delta \phi_2)$ of two arms with respect to the third one (reference). These are thermo-optic phase shifts that can be tuned by  means of ohmic resistors. When a set of voltages $(V_1, V_2)$ is applied to the resistors, a different global phase shift is generated along each optical path, thus changing the action of the device. A scheme of the device is reported in Fig.~\ref{concim}. The chip is studied in the single-photon regime. Pairs of photons with wavelength $785$nm are generated by spontaneous parametric down conversion process in a

barium borate (BBO) crystal. A single photon of the pair is coupled into a fiber array connected to one of the circuit's inputs, while the other  photon acts as a trigger. Single-photon detectors are placed at the output fibers of the device and coincidence events between the trigger and the photon injected inside the chip are recorded. In this way, single-photon probabilities $P(i \rightarrow j)$ for each output $j$ ($j=1,2,3$) are measured by changing the input arm $i$ of the single-photon state and tuning the power dissipated on the internal resistors. In particular, we tune the voltages applied to each resistor independently, while  keeping the others off.
%set 50 different tension values on each resistor, obtaining output probabilities at fixed input for 50x50 different points.
Due to the resistance dependence on its temperature and due to the thermal crosstalk between resistances, a given pair of  voltages, applied on the thermo-resistors $R_1$ and $R_2$ respectively, implies a variation of the two phases described by the following approximate response function containing linear and quadratic dependencies from the dissipated power:
\begin{equation}
    \Delta\phi_{i}=\sum_{j=1}^2 \left(\alpha_{ij}P_{R_j}+\alpha_{ij}^{NL}P_{R_j}^{2} \right),
    \label{fasitensioni}
\end{equation}
where the dissipated power is $P_{R_j}=V_j^2/R_j$ using Ohmic approximation for the two resistors, while $\alpha_{ij}$ and $\alpha_{ij}^{NL}$ are, respectively, the linear and non-linear response coefficients associated to the phase shift $\Delta\phi_{i}\,(i=1,2)$ when dissipating power on resistor $R_j$.

The two voltages $V_1$ and $V_2$ are thus directly mapped into the two physical relative phase-shifts $\Delta \phi_1$ and $\Delta \phi_2$. In standard characterisation, a theoretical model of the circuit allows to recover the likelihood describing the output probabilities through a fit of the measured probabilities. This, in turn, allows to extrapolate dynamic and static parameters of the chip \cite{polino2019experimental,valeri2020experimental}. The aim of this work is to avoid relying on knowledge about both the theoretical model of circuit and  the response function in Eq. \eqref{fasitensioni}. In fact, this would prove inefficient for the characterisation of mass-produced devices. The goal, instead, is to generate a mapping between voltages and output probabilities using only a limited set of measurements. We exploit the NN approach exactly to realize such goal (Fig. \ref{concim}).

%\textcolor{green}{Questa frase sembra ambigua perche dopo un paio di frasi diciamo che usiamo la maximum likelihood. Quindi in qualche modo usiamo la likelihood.. Pero la NN in principio non ne avrebbe bisogno e l'approccio non richiede la conoscenza della likelihood, giusto?.. Quindi o cambierei la frase della maximum likelihood magari dicendo che serve solo come primo step di principio non necessario oppure questa frase la leverei...} 

\begin{figure*}[htb!]
\begin{center}
{\includegraphics[width=\textwidth]{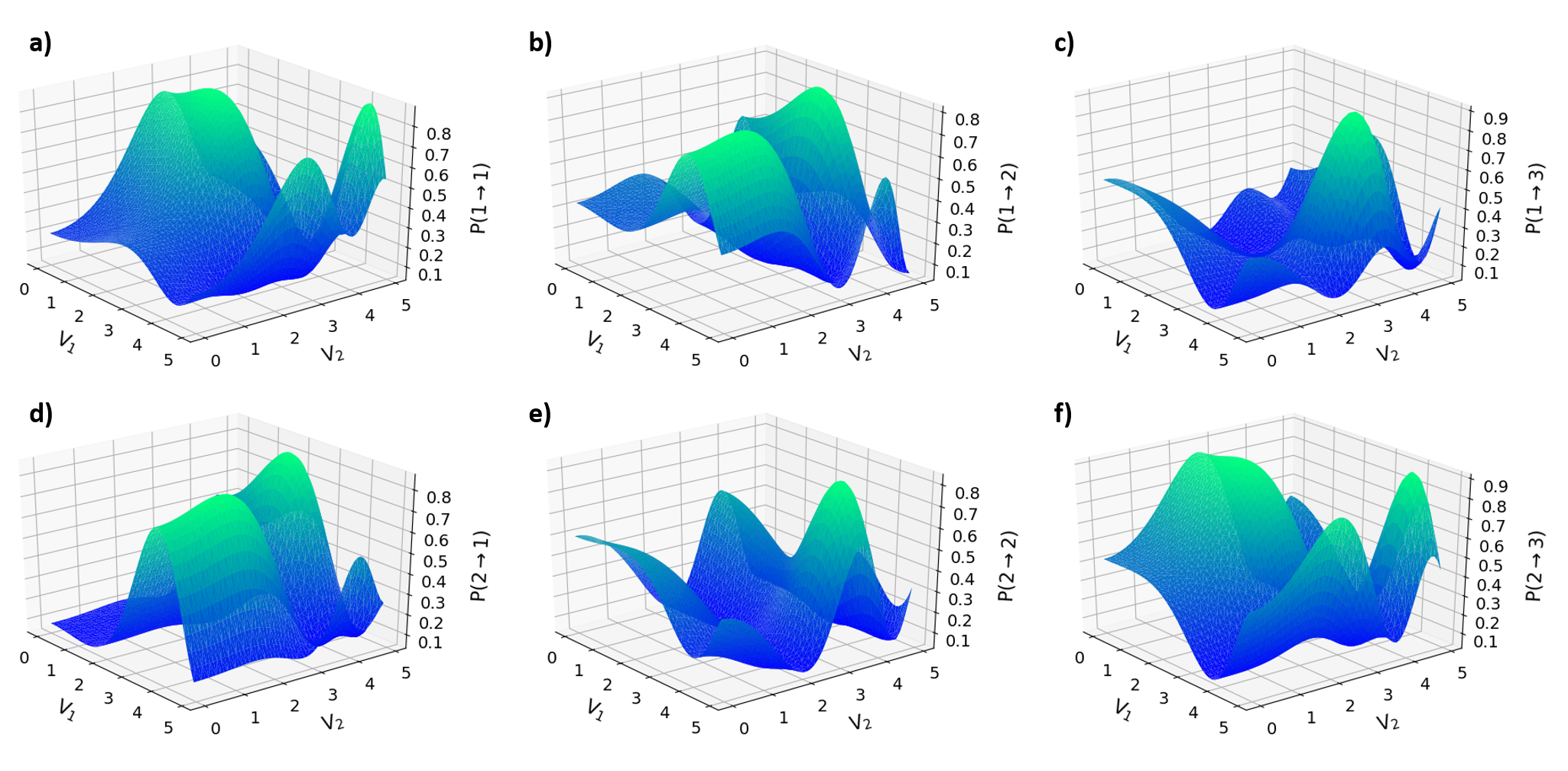}}
\caption{
Examples of input-output probabilities $P(i \rightarrow j)$ for each input $i$ and output $j$ ($i,j=1,2,3$) of the chip, resulting from predictions of a NN trained with simulated data, as function of the applied voltages $V_1$,$V_2$ over two ohmic resistors. \textbf{a)-c)} Output probabilities $P(1 \rightarrow j)$ obtained by injecting single photons along input 1. \textbf{d)-f)} Output probabilities  $P(2 \rightarrow j)$ obtained by injecting single photons along input 2.}
\label{img2}
\end{center}
\end{figure*}

\subsection{Neural network architecture and performances}

We start using simulated data to inspect the algorithm requirements both in terms of network architecture and amount of training data needed to obtain good performances when evaluating new examples. This preliminary step has the only purpose of identifying more efficiently the structure of the NN for the calibration of a general three-arm interferometer. Therefore, a likelihood function describing exactly the device is not needed; the simulated data represent roughly its typical approximate likelihood. In our case   the 
simulated data are obtained based on single-photon input-output probabilities $P(i\rightarrow j)$ estimated from experimental measurements through a maximum-likelihood technique. To train the NN, data are divided in a vector of input features $\vec{x}$ corresponding to the $9$ input-output probabilities obtained when applying a given pair of controlled voltages, which constitute the elements of the output vector $\vec{y}$:

\begin{equation}
\big\{\vec{x},\vec{y}\big\}_k = \big\{P(i\rightarrow j), (V_1,V_2)\big\}_k \text{ for } i,j = 1,2,3.
\end{equation}
The index $k = 1,\dots,N$ refers to each training example and $N$ represents the total number of training data.  The vector of probabilities $\vec x$ is constructed using counts extracted from a Poisson distribution whose mean values correspond to those assessed from the maximum-likelihood reconstruction. This procedure is needed for the NN to properly account for the presence of such source of uncertainty. A part of the data, viz. 15\% of the whole set, is used as a validation set: it is not directly employed for the training, but rather to obtain an independent estimate on the training error. This is necessary to avoid overfitting.

We have tested different NN architectures with a different number of hidden layers and neurons per layer, studying the network performances for different activation functions, initialization parameters and optimization algorithms. Finally,  we have chosen the one which obtains the smallest Root Mean Squared Error (RMSE) on the validation set. We notice that the full set of $9$ probabilities is redundant for the network training and satisfactory results can indeed be achieved training the network with only $6$ of them, obtained when injecting a photon respectively in the first and second input of the device.
%We notice that, since the probabilities are not independent, the full set of $9$ probabilities is redundant for the network training and satisfactory results can indeed be achieved training the network with only $6$ of them.
The training consists in tuning the model's parameters to minimize the RMSE associated to each example of the training set; for this purpose the gradient of the loss function, corresponding to the summation of the RMSE over all the training examples, with respect to all the network's parameters is computed using the backpropagation method \cite{DeepLearning}. In the next step the gradient is used to minimize the loss function using the optimization algorithm Adam \cite{Adam}.

We train the network with the results obtained after the application of $53$ different tensions values to each of the two resistors in the device. This gives a tension grid with $53 \times 53 = 2809$ different tension pairs associated to the relative input-output probabilities available to train the network. Therefore, each example in the training set, consists in the tensions pairs associated to $6$ probability values. The trends of the values predicted by the trained-NN are reported in Fig. \ref{img2} %DOMANDA Mauro: le 6 probabilita indipendenti non dovrebbero essere due output associati a ogni input? Invece nella fig.2 facciamo vedere tutte e tre le probabilità di output associate ai due input..
as a function of the applied voltages on the two resistors. However, to obtain a good estimation in the full range of accessible tensions, it was necessary to incorporate into each training example the further set of probabilities $\tilde{P}(i\rightarrow j)$, to which we refer as kicks, as follows:
\begin{equation}
\begin{split}
&\big\{\vec{x},\vec{y}\big\}_k =\\
&\big\{(P(i\rightarrow j),\tilde{P}(i\rightarrow j)), (V_1,V_2,V_1+\Delta V_1,V_2+\Delta V_2)\big\}_k 
\end{split}
\end{equation}
where the values $\tilde{P}(i\rightarrow j)$ are added by considering the probabilities obtained by changing $V_1\,(V_2)$ of a fixed value $\Delta V_1\,(\Delta V_2)$ and the length of the $\vec{x}$ and $\vec{y}$ vectors is $K$. This is necessary to remove ambiguities in the evaluation of the overall function, providing additional information to the network. More specifically, such requirement is due to the non-monotonicity of the output probabilities, resulting in the presence of multiple parameter points that correspond to the same probability values. Thus, in absence of additional data sets it is not possible to distinguish between those points.
Comparing the network results on the validation set, the RMSE improves by $85\%$ thanks to the additional information provided with the tension kick, when evaluated in the full range of tension values. The advantage obtained is independent from the specific value of $\Delta V_1$ and $\Delta V_2$, as long as they are large enough to give information about different regions of the inspected functions. Enlarging the data in each training example requires to increase the number of nodes in the input and output layers of the network. The architecture that allows to achieve the best performances, among the ones considered, is a network with $12$ input nodes and $4$ output ones separated by $3$ hidden layers with $200$ nodes each (Fig.~\ref{concim}). All the nodes, except the output ones which are activated by a linear function, are activated by a rectified linear unit function initializing their weights with random values extracted from a normal distribution centred in zero and with variance $\sigma^2 = 2/n$, where $n$ is the number of neurons in the previous layer. 

We stop the training  after $250$ epochs since the loss function on the validation set stops decreasing. To analyze the variability among different trainings, we study the results obtained, starting from the same datasets, after performing $50$ independent trainings of the network. The mean value of the Normalized Root Mean Squared Error (NRMSE) is given by: 

\begin{equation}
\text{NRMSE}=\frac{1}{\sqrt{K}}\frac{||\vec{y}-\vec{\hat y}||}{y_{max}-y_{min}}
\end{equation}
where $||\cdot||$ indicates the Euclidean norm. It is calculated on the validation set in such configuration obtaining a value of $\text{NRMSE}=0.015\pm0.001$. 
After the network has been trained, its performances have been evaluated on an independent test set of $100$ different examples selected randomly among the possible tensions pairs of the $53 \times 53$, adding Poissonian noise on the detection events determining the probabilities values. Notably, since both the training and the test data include by random Poissonian noise, the resulting NN is robust when evaluating new noisy examples. To quantify how close the network estimation of the tension values is to the true ones, we evaluate the cosine similarity between the vector of network outputs $\vec{y}$, corresponding to the $4$ tensions for all the test examples, and the expected results $\vec{\hat y}$, as follows:

\begin{equation}
c=\frac{\vec{y}\cdot\vec{\hat y}}{||\vec{y}||\cdot||\vec{\hat y}||},
\end{equation}
 which is equal to $1$, when the prediction $\vec{\hat y}$ is equal to the true value $\vec{y}$. Moreover, to estimate how much the cosine similarity depends on the random sample selected, we compute its value on $500$ repetitions each containing $100$ examples extracted randomly among the available ones, obtaining a value of $c = 0.999\pm0.001$.

\begin{figure}[h]
\begin{center}
{\includegraphics[width=\columnwidth]{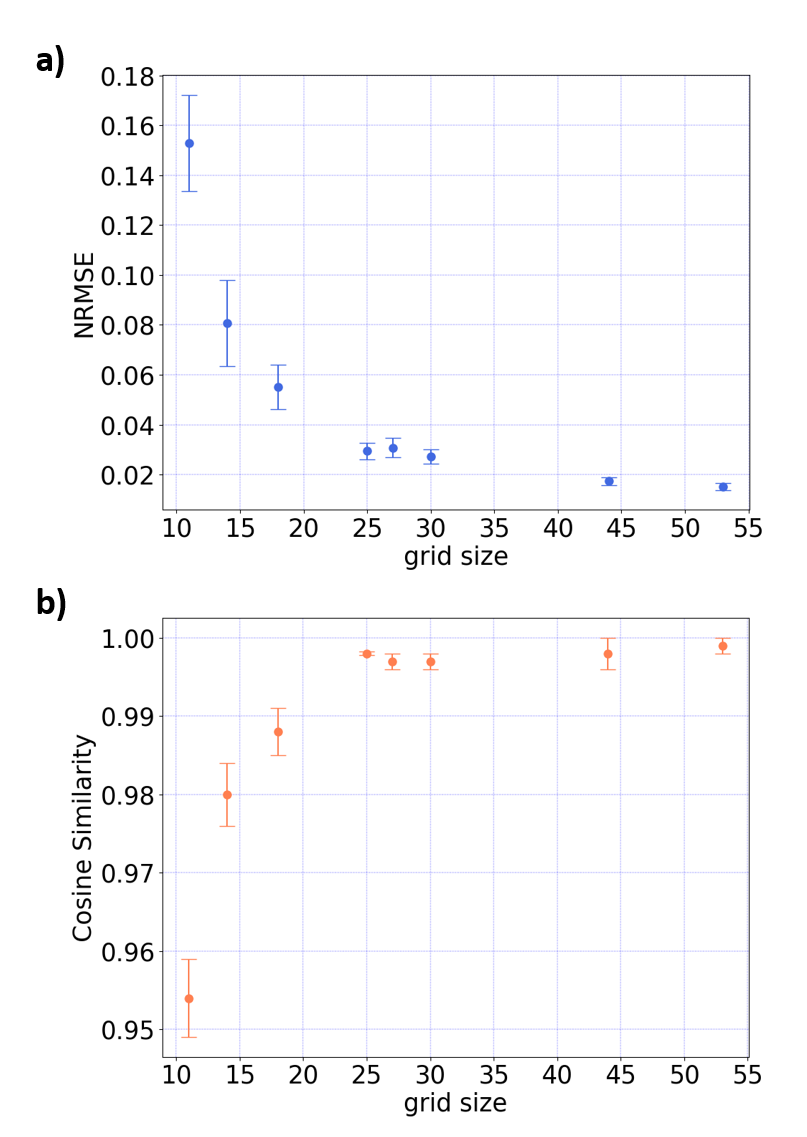}}
\caption{NN training with different amount of data grid size. Training performances are shown in terms of \textbf{a)} NRMSE computed over the validation set and \textbf{b)} cosine similarity on the test one.}
\label{img1}
\end{center}
\end{figure}

Once established the network architecture, we investigate how the NRMSE on the validation set and the cosine similarity between the network estimation and the true tensions values change reducing the number of tension pairs used for the training. For all the different configurations the size of the training set, and consequently of the validation one, changes depending on the dimension of the tension grid used. Conversely, the number of examples making up the test set remains fixed to $100$. For all the new training configurations the test data are still extracted randomly from the largest grid. This choice is performed to assess how much reducing the data for the training affects the final network estimation of new examples.  Figure \ref{img1} reports the NRMSE on the validation set, obtained from multiple trainings of the network and the cosine similarity among the network estimation and the expected values in the test set. As expected, the NRMSE achieved by the network decreases as the number of training examples increases, allowing a better reconstruction of the function mapping the input vector onto the output one. A better reconstruction of this function grants higher network performances on the independent test set, as shown by the growth of the cosine similarity between the reconstructed tensions vectors by the NN and the real one. The error on the cosine similarity values gives an indication about the variability linked to the analysis of different examples that randomly fall in different regions of the probabilities functions. In parallel, the error on the NRMSE values depends on the results of different trainings of the algorithm that, starting with random initial parameters, can end up in slightly different conditions.

\begin{figure}[h!]
\begin{center}
{\includegraphics[width=\columnwidth]{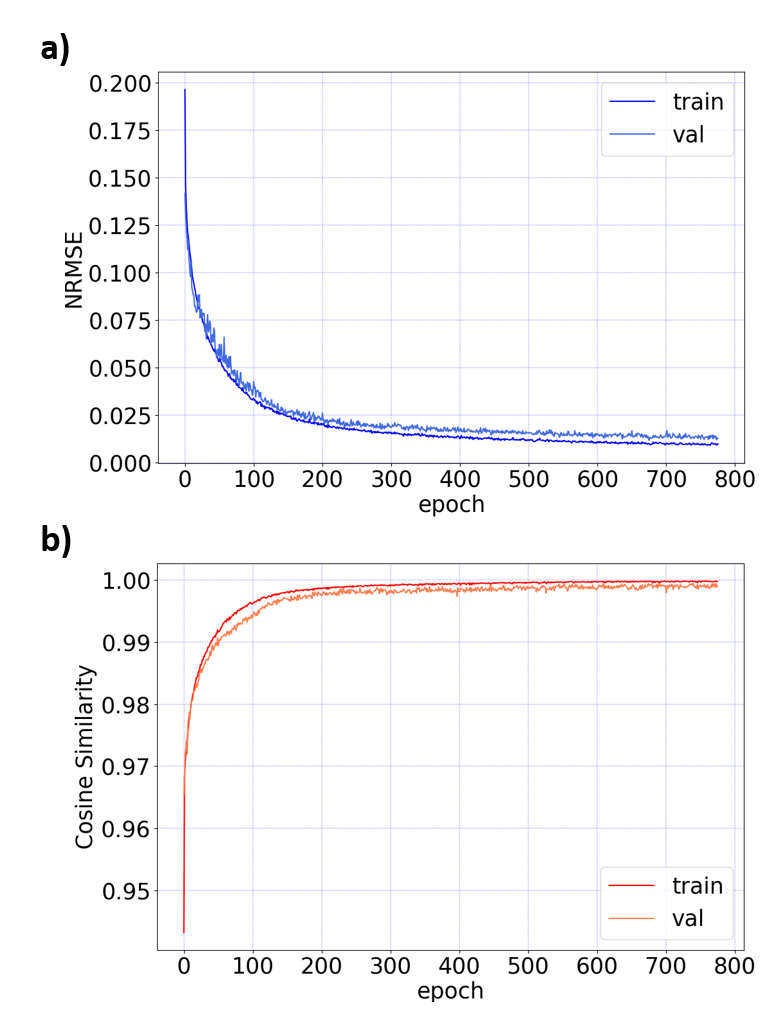}}
\caption{Study of NN performances by using experimental dataset for both training and validation stages. \textbf{a)} NRMSE and \textbf{b)} cosine similarity are shown as function of the computed epochs.}
\label{img4}
\end{center}
\end{figure}

\begin{figure*}[htb!]
\begin{center}
{\includegraphics[width=\textwidth]{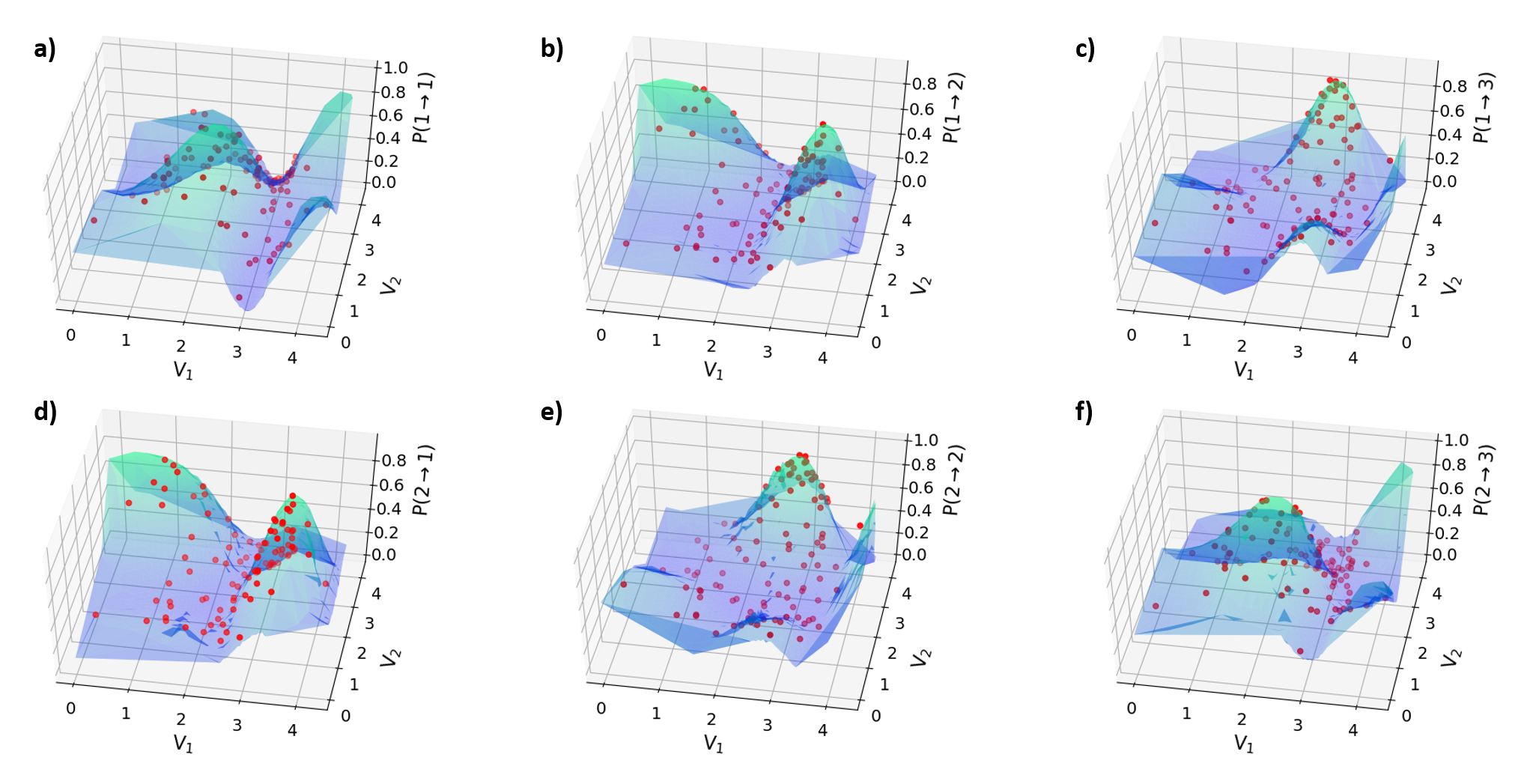}}
\caption{Comparison between voltages predicted from NN (red dots) and grid of experimental values (blue surface) for different single-photon input-output probabilities.}
\label{img3}
\end{center}
\end{figure*}

\subsection{Results on experimental data}

Here, the network architecture resulting from the previous analysis has been implemented and tested training the NN directly on actual experimental data. This is possible thanks to the network efficiency to learn the function which maps the input-output probabilities to the tensions applied. In this way, a good estimate of the tensions values is obtained. %\textcolor{red}{, in principle allowing to derive the optical phases of the interferometer when new detection events are registered, without the need of an explicit model of the device.} \textcolor{blue}{. In principle, by knowing only the response of the optical phases as function of the applied voltages, one could, using this approach, estimate the phases without the need of an explicit model of the whole device.} 
If the explicit model Eq. (1) is available, this same procedure is also effective to calibrate the response of the device explicitly in terms of the phases; the exact description of the further propagation and measurement steps is not needed.
 We use the data obtained from a $50 \times 50$ tensions grid applied on a second integrated device with the same layout, to train the same network as the one described above. The training with such data takes longer to reach the region where the loss function on the validation set stops decreasing. This is shown  in Fig. \ref{img4}, where the NRMSE and the cosine similarity are reported as a function of the training epochs. The network tensions estimations when $100$ new data are acquired are reported in Fig. \ref{img3}, showing the ability of the network to make accurate predictions of the applied voltages.

\section{Conclusions}
\label{conclusions}

We have reported on the application of a NN based algorithm to perform the calibration of integrated devices  depending on two parameters.  In this investigation we relied on knowledge of a model to identify the most appropriate regime for collecting the training set. However, this is by no means a necessary step in that the use of the NN itself incorporates the same information that would be present in the model. Remarkably, the NN is able to account for spurious effects such as, in our case, cross-talk between thermal actuators, which are otherwise intricate to describe. It can be foreseen however that some basic modelling of the device could be nevertheless beneficial. The successful characterization of two devices based on a single approximate model shows that the NN performance does not heavily depend on the model's level of detail. In the same vein, some anticipation of the device output, might reveal whether ambiguities may be present in the chosen range of parameters. We have shown that this is easily accounted for by introducing additional data as input to the NN. 

This study brings forward machine learning applications in two respects: it goes beyond optimization of the employed resources when these are severely constrained and shows that this characterization method extends beyond the single parameter regime. The obtained results give evidence that NN can provide an effective, robust and reliable tool for the calibration of complex sensors that depend on multiple parameters, with the advantage of requiring no detailed model of their internal operation. %\textcolor{green}{Notably, the algorithm, being able to work with noisy experimental data, is demonstrated to be a robust approach for the characterization of real sensors. }

\begin{acknowledgments}
This work is supported by the European Research Council (ERC) Advanced Grant CAPABLE (Composite integrated photonic platform by femtosecond laser micro-machining, Grant Agreement No. 742745), and from the European Union's Horizon 2020 research and innovation programme under the PHOQUSING project GA no. 899544. I.G. is supported by Ministero dell'Istruzione, dell'Universit\`a e della Ricerca Grant of Excellence Departments (ARTICOLO 1, COMMI 314-337 LEGGE 232/2016).
\end{acknowledgments}

\bibliography{refs_NN.bib}
\bibliographystyle{apsrev4-2}

\end{document}